\documentstyle[epsf,epsfig,aps]{revtex}
\draft
\input epsf
\begin{document}
\twocolumn[\hsize\textwidth\columnwidth\hsize\csname@twocolumnfalse%
\endcsname
\title{Quantum voltage oscillations observed on segments of an inhomogeneous superconducting loop}
\author{S. \ V. \ Dubonos, \ V. \ I. \ Kuznetsov, and \ A. \ V. \ Nikulov}

\address{Institute of Microelectronics Technology and High Purity Materials, Russian Academy of Sciences, 142432 Chernogolovka, Moscow District, RUSSIA}

\maketitle
\begin{abstract}
{The theoretical prediction published in {\it Phys.Rev. B} {\bf 64}, 012505 (2001) is
verified. In accordance with this prediction a dc voltage oscillating with magnetic
field is observed on segments of an inhomogeneous loop in the temperature region
close to the superconducting transition. The temperature dependence of the amplitude
of the voltage oscillation implies that a transformation of the energy of an external electrical
noise to the dc power is observed.}
 \end{abstract}  ]
 \narrowtext

According to the universally recognized explanation  \cite{tink75}
of the Little-Parks  (LP) experiment \cite{little}
the resistance oscillations are observed  \cite{repeat} because of the fluxoid
quantization \cite{little,tinkham}. Because of the quantization
$$\oint_{l}dl v_{s} = \frac{\pi \hbar}{m} (n -\frac{\Phi}{\Phi_{0}}) \eqno{(1)} $$
the circulation of the velocity of superconducting pairs $v_{s}$ can not be
equal zero when the magnetic flux  $ \Phi$ contained within a loop is not
divisible by the flux quantum $\Phi_{0} = \pi \hbar c/e$  \cite{tink75}. Therefore the
energy of superconducting state increases and as consequence the  $T_{c}$
decreases  when  $ \Phi \neq n\Phi_{0}$, $\Delta  T_{c} \propto -v_{s}^{2}
\propto  -(n -\Phi/\Phi_{0})^{2}$ \cite{tink75}. The resistance increases at  $\Phi \neq
n\Phi_{0}$ because of the $T_{c}$ decrease:  $\Delta R =-
(dR(T-T_{c})/dT)\Delta T_{c} \propto (dR/dT)v_{s}^{2}$ \cite{tink75}.

Without any external current the  $v_{s}$ value is proportional to the
superconducting screening current $v_{s} \propto I_{sc} = sj_{sc} = s2en_{s}v_{s} =
s2e <n_{s}^{-1}>^{-1}(\pi \hbar/ml) (n -\Phi/\Phi_{0})$. The spatial average $<n_{s}^{-1}> =
l^{-1}\oint_{l}dl n_{s}^{-1}$ of the value $n_{s}^{-1}$ inverse of the density of
superconducting pairs $n_{s}$ is handy to use because the
$I_{sc}$ value should be constant along the loop in the
stationary state. The magnetic flux $LI_{sc}$
induced by the screening current  $I_{sc}$ is small
$LI_{sc} \ll \Phi_{0}$  at  $T \simeq T_{c}$  (when  the $n_{s}$ value is small) and therefore  $\Phi = BS
+ LI_{sc}\simeq BS$ \cite{tink75}. Here $B$ is the magnetic induction
induced by an external magnet; $S$ is the area of the loop.

Because of the thermal fluctuation the $<n_{s}^{-1}>^{-1}$ and $I_{sc}$ values
oscillate strongly in time at $T \approx T_{c}$. The $n$ can be also an random integer number
but in the majority of cases this integer number is corresponded  to minimum possible
value $v_{s}^{2} \propto (n -\Phi/\Phi_{0})^{2}$. Therefore the time average
$\overline{(n -\Phi/\Phi_{0})} = t_{long}^{-1} \int_{t_{long}}dt  (n -\Phi/\Phi_{0})
\simeq  (n - \Phi/\Phi_{0})_{min}$ when $\Phi$ is not close to  $(n+0.5) \Phi_{0}$. This
is corroborated by the comparison of the theoretical dependence
 $\Delta  T_{c} \propto -(n -\Phi/\Phi_{0})_{min}^{2}$ \cite{tink75}
with the experimental data for $\Delta  T_{c}(\Phi)$ (see for example Fig.4 in
\cite{repeat}).

Thus, according to the LP experiment the persistent current
$$I_{p.c.} = \overline{I_{sc}} \approx s2e\overline{<n_{s}^{-1}>^{-1}}
\frac{\pi \hbar}{ml}(n -\frac{\Phi}{\Phi_{0}})_{min} \eqno{(2)} $$
i.e. a current with a direct component
$ \overline{I_{sc}} =  t_{long}^{-1} \int_{t_{long}}dt I_{sc}$ flows along the loop at a constant
magnetic flux, $ \Phi \neq n\Phi_{0}$ and $ \Phi \neq (n+0.5) \Phi_{0}$,
and a nonzero resistance along the loop $R_{l} > 0$  in spite of the Ohm's law
$R_{l}I_{sc} = \oint_{l}dl E = -(1/c)d\Phi/dt$.

In order to explain the existance of the persistent
current at zero Faraday's voltage and nonzero resistance the quantum force is
proposed to introduce in the paper \cite{QuaForce}. It is predicted also in
\cite{QuaForce} that not only the persistent current but also a persistent voltage
can be observed on segments of an inhomogeneous superconducting loop.
The value and sign of the persistent voltage as well as of the persistent current should
depend in a periodic way on a magnetic flux $\Phi$.
Such voltage oscillations $V_{os}(\Phi/\Phi_{0})$ without any external current
was predicted first in \cite{jltp98}. These oscillations can be observed in the
temperature region close to $T_{c}$ \cite{QuaForce,jltp98}.

The possibility of the persistent voltage predicted in \cite{QuaForce} is obvious
from the analogy with a conventional inhomogeneous loop with a current $I_{sc}$
induced by Faraday's voltage $I_{sc} = R_{l}^{-1}\oint_{l}dl E = -(1/ R_{l}c)d\Phi/dt$.
According to the Ohm's law $\rho j_{sc} = E = -\bigtriangledown V - (1/c)dA/dt =
 -\bigtriangledown V - (1/cl)d\Phi/dt$ the potential difference
$$V = (<\rho>_{l_{s}} - <\rho>_{l})l_{s}j_{sc} \eqno{(3)}$$
should be observed on a segment $l_{s}$ of an inhomogeneous
loop at $j_{sc} \neq 0$ if the average resistivity along this segment
$<\rho>_{l_{s}} = \int_{l_{s}} dl \rho /l_{s}$ differs from the one along
the loop $<\rho>_{l} = \oint_{l} dl \rho /l$.

The object of the present work is an experimental verification of the
theoretical prediction \cite{QuaForce} and of the analogy with a
conventional loop. According to both the prediction \cite{QuaForce}
and the analogy (3) the voltage oscillations $V_{os}(\Phi/\Phi_{0})$
without any external current can be observed  in a inhomogeneous
loop where $<\rho>_{l_{s}} - <\rho>_{l} \neq 0$ and should not
observed in a homogeneous one where $<\rho>_{l_{s}} - <\rho>_{l} = 0$.
In order to investigate the influence of the heterogeneity of loop segments
we made both symmetrical and asymmetric loops in each investigated
structure (see Fig.1). Because of the additional potential contacts
$V_{3}$ the higher and lower segments of the lower loop (on Fig.1)
can have a different resistance at $T \simeq T_{c}$ when $\Phi \neq n\Phi_{0}$,
whereas the upper and lower segments of the higher loop should have
the same resistance if any accidental heterogeneity is absent.

We used the mesoscopic Al
structures, one of them is shown on Fig.1. These microstructures are
prepared using an electron lithograph developed on the basis of a
JEOL-840A electron scanning microscop. An electron beam of the lithograph
was controlled by a PC, equipped with a software package for proximity
effect correction "PROXY". The exposition was made at 25 kV and 30 pA. The
resist was developed in MIBK: IPA = 1: 5, followed by the thermal
deposition of a high-purity Al film 60 nm and lift-off in acetone. The
substrates are Si wafers.  The measurements are performed in a standard
helium-4 cryostat allowing us to vary the temperature down to 1.22 K. The
applied perpendicular magnetic field, which is produced by a superconducting coil, never
exceeded 35 Oe. The voltage variations down to 0.05 $\mu V$ could be
detected. In order to diminish an influence of an external electric noise
resistances were used as cold filters.

\begin{figure}[bhb] \vspace{0.1cm}\hspace{-1.5cm}
\vbox{\hfil\epsfig{figure= 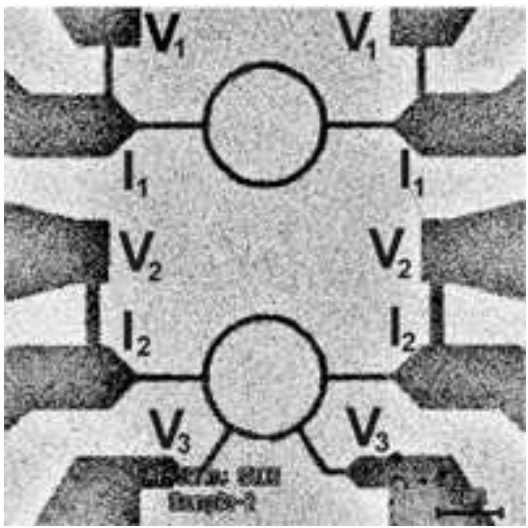,width=7cm,angle=0}\hfil}
\vspace{0.75cm} \caption{An electron micrograph one of the aluminum loop
samples.  $I_{1}$ and $V_{1}$ are the current and potential contacts of
the  symmetrical loop. $I_{2}$ and $V_{2}$ are the current and potential
contacts of the asymmetric loop. $V_{3}$ are the additional potential
contacts of the asymmetric loop.} \label{fig-1} \end{figure}

We have investigated the dependencies of the dc voltage $V$ on the magnetic
flux $\Phi \approx BS$ of some round loops with a diameter 2r = 1, 2 and 4
$\mu m$ and a linewidth w = 0.2 and 0.4 $\mu m$ at the dc measuring current
$I_{m}$ and different temperature close to $T_{c}$. The sheet resistance
of the loops was equal approximately $0.5 \
\Omega /\diamond $ at 4.2 K, the resistance ratio $R(300 K)/R(4.2 K)
\approx 2$ and the midpoint of the superconducting resistive transition
$T_{c} \approx  1.24 \ K$. All loops exhibited the anomalous features of
the resistive dependencies on temperature and magnetic field which was before
observed on mesoscopic Al structures in some works \cite{repeat,anomal}.

The results of our measurements Fig.2-4 corroborate the theoretical
prediction \cite{QuaForce} and the analogy with a conventional loop (3).
We observe the conventional LP oscillations of the resistance  \cite{little} at the
symmetrical loops Fig.2. This result repeats the observations made before
in  many works and is not new result. In accordance with the prediction
\cite{QuaForce} and the analogy with a conventional loop (3) the voltage
measured at the $V_{1}$ contacts equals zero at the measuring current
$I_{m} = 0$ and the oscillations with the amplitude
$\Delta V \approx \Delta R_{m}(\Phi/\Phi_{0},T/T_{c}(I_{m}))I_{m}$
are observed only at $I_{m} \neq 0$ Fig.2. The LP oscillations are observed
against a background of an anomalous
behaviour, the downfall before the disappearance of
the LP oscillation Fig.2.  Such anomalies were observed also on other our loops and in
other works \cite{anomal}.

\begin{figure}[bhb] \vspace{0.1cm}\hspace{-1.5cm}
\vbox{\hfil\epsfig{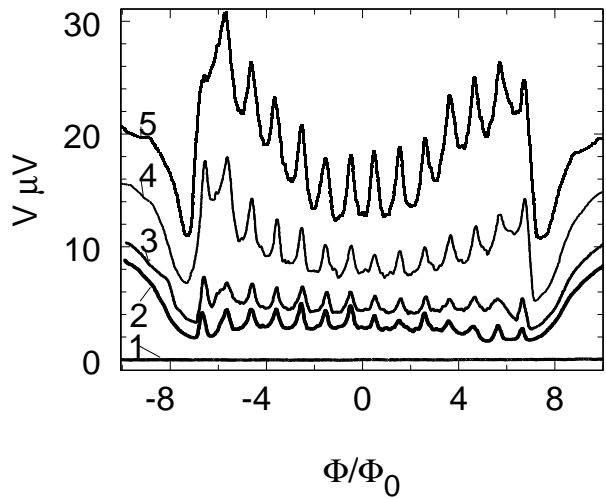}\hfil}
\vspace{0.75cm} \caption{The voltage oscillations measured on the $V_{1}$
contacts of the symmetrical loop with 2r = 4 $\mu m$ and w = 0.2 $\mu m$ at
different $I_{m}$ values between the  $I_{1}$ contacts: 1 - $I_{m} = 0.000
\ \mu A$; 2 - $I_{m} = 1.83 \ \mu A$;  3 - $I_{m} = 2.10 \ \mu A$; 4 -
$I_{m} = 2.66 \ \mu A$; 5 - $I_{m} = 3.01 \ \mu A$. $T = 1.231 K$ is
corresponded to the bottom of the resistive transition } \label{fig-1}
\end{figure}

In contrast to the symmetrical loops we observe new phenomenon, which
was not published before anywhere, at the voltage measurement on the
contacts $V_{2}$ and $V_{3}$ of the asymmetric loop. We observe no resistance
but voltage oscillations: $V \approx  V_{os}(\Phi/\Phi_{0}) + R_{nos}I_{m}$
Figs.3,4. The greatest  amplitude $\Delta V$ of the voltage oscillations are
observed at  $I_{m} = 0$ and the  $\Delta V$ value does not increase
with the $I_{m}$ Fig.3 in contrast to the symmetrical loop Fig.2.
The voltage oscillations at a high  $I_{m}$ are observed against
a background of an anomalous behaviour, the negative
magnetoresistance $R_{nos}$ Fig.3, like qualitatively the one
observed in the symmetrical loop Fig.2.

The voltage oscillations  $V_{os} (\Phi/\Phi_{0})$ observed at $I_{m} = 0$ Figs.3,4
correspond with the LP experiment (2) and the analogy with a
conventional loop (3). According to (2) and (3)
$$V_{os} (\frac{\Phi}{\Phi_{0}}) = (<\rho>_{l_{s}} - <\rho>_{l})l_{s}j_{p.c.} (\frac{\Phi}{\Phi_{0}}) \eqno{(4)} $$
should be observed when loop segments have different resistivity
$<\rho>_{l_{s}} \neq <\rho>_{l}$. The relation (4) describes
enough well the voltage oscillations observed at  $I_{m} = 0$ Figs.3,4.
The $V_{os} (\Phi/\Phi_{0})$ oscillations Figs.3,4 and the LP oscillations Fig.2 have
the same period and  are observed in the same temperature region
near $T_{c}$ where $j_{p.c.} \neq 0$ and $<\rho>_{l}= R_{l}/l  > 0$.
The magnetic field regions, where they
are observed, are also close. The oscillations on Fig.2 are observed
in more wide magnetic field region than on Figs.3,4 because the width
of the wire defining the loop in the first case w = 0.2 $\mu m$ is
smaller than in the second case w = 0.4 $\mu m$. In any real case
only some oscillations are observed because a high
magnetic field breaks down the superconductivity, i.e $I_{p.c.}$, in the wire
defining the loop. According to (1) $v_{s} = (\pi
\hbar /m) Br/2$ along the loop and $v_{s} = (\pi \hbar /m) Bw/2$ along the
boundaries of the wire at $n=0$. Therefore a limited number of oscillations
$\propto 2r/w$ are observed.

\begin{figure}[bhb]
\vspace{0.1cm}\hspace{-1.5cm}
\vbox{\hfil\epsfig{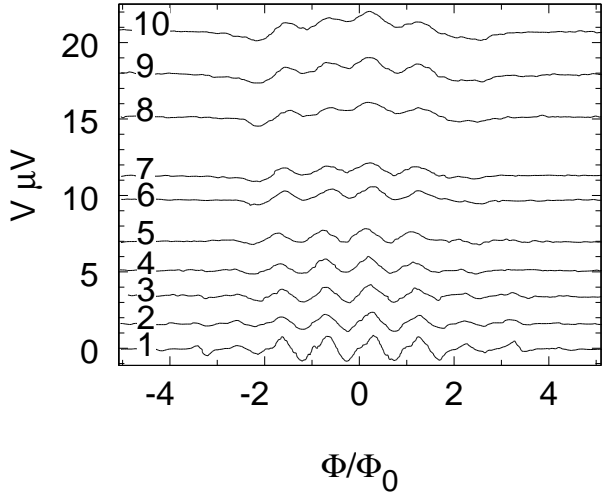}\hfil}
\vspace{0.75cm} \caption{The voltage oscillation measured on the $V_{2}$
contacts of the asymmetric loop with 2r = 4 $\mu m$ and w = 0.4 $\mu m$ at
different value of the measuring current between the  $I_{2}$ contacts: 1 -
$I_{m} = 0.000 \ \mu A$; 2 - $I_{m} = 0.29 \ \mu A$;  3 - $I_{m} = 0.65 \
\mu A$; 4 - $I_{m} = 0.93 \ \mu A$;  5 - $I_{m} = 1.29 \ \mu A$;  6 -
$I_{m} =1.79 \ \mu A$;  7 - $I_{m} = 2.06 \ \mu A$;  8 - $I_{m} = 2.82 \
\mu A$;  9 - $I_{m} = 3.34 \ \mu A$; 10 - $I_{m} =3.85 \ \mu A$.  $T =
1.231 K$ is corresponded to the bottom of the resistive transition }
\label{fig-1} \end{figure}

According to (4) the amplitude of the voltage oscillations $\Delta V $ is
proportional to the lengths of the segment $l_{s}$. The $\Delta V $ values
observed between $V_{3}$ and between $V_{2}$ differ approximately in 6 times Fig.4
whereas the $l_{s}$ values between these contacts differ in 3 times only.
This means that the $(<\overline{\rho}>_{l_{s}} -
<\overline{\rho}>_{l})$ value in the first case is smaller than in the second.
Such difference ought be expected because both additional and main contacts
should influence on superconducting state of loop segments.

Using the analogy with a conventional loop (4) we can evaluate
the lower limit of the persistent current $I_{p.c.} = sj_{p.c.}$ inducing
the voltage oscillations with the amplitude $\Delta V \approx 1 \ \mu V$
observed on the $V_{2}$ contacts  Figs.3,4. According to (4)  $V_{os} =
0.5(R_{hs} - R_{ls})I_{p.c.}$ on these contacts, where  $R_{hs}$ and
$R_{ls}$ are the resistances of the higher and  lower  segments at
the temperature of measurement. Because $0 < R_{hs},R_{ls} < R_{sn}$
the $I_{p.c.}$ value should exceed $2V_{os}/R_{sn}$, where $R_{sn}$ is
the resistance of the higher and  lower segments in the normal state.
The oscillations Figs.3,4 are observed on segment with $R_{sn} \simeq  5 \ \Omega $.
Consequently the amplitude of the persistent current (2) inducing these
oscillations $\Delta I_{p.c.} \geq 0.4 \ \mu A$.

\begin{figure}[bhb]
\vspace{0.1cm}\hspace{-1.5cm}
\vbox{\hfil\epsfig{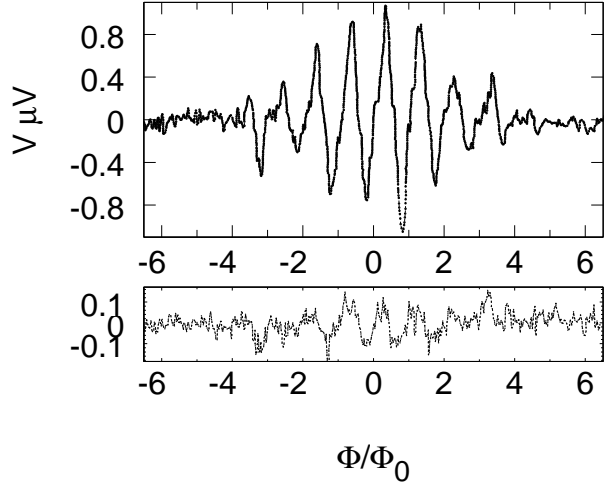}\hfil}
\vspace{0.75cm}
\caption{Oscillation of the voltage measured on the $V_{2}$ contacts (upper
curve) and on the $V_{3}$ contacts  (lower curve) of the asymmetric loop
with 2r = 4 $\mu m$ and w = 0.4 $\mu m$. $I_{m} = 0$. $T = 1.231 K$
corresponded to the bottom of the resistive transition. } \label{fig-1}
\end{figure}

We can also evaluate in order of value the persistent current inducing the
LP oscillations Fig.2. The persistent
current at $\Phi = (n+0.5) \Phi_{0}$ induces a resistance change which
equals in order of value the one induced by an increase of the measuring
current $\Delta I_{m} \approx 0.5 \ \mu A$ Fig.2. One should assume
that $\Delta R_{m}/\Delta I_{m}$ and $\Delta R_{m}/\Delta I_{p.c.}$
are close in order of value. According to this estimation the persistent current inducing
the LP oscillations Fig.2 can induce the voltage oscillations at $ I_{m} = 0$ Figs.3,4.

Thus,  the existance of the persistent current (2) and the analogy with a
conventional loop (3) allow to explain enough well the voltage oscillations
observed at $I_{m} = 0$ Figs.3,4. But this does not mean that our result
shown on Figs.3,4 is not new in essence. There is an important difference
between the conventional current $I_{sc} = R_{l}(-1/c)d\Phi/dt$ induced by
the Faraday's voltage $(-1/c)d\Phi/dt$ and the persistent current at $d\Phi/dt = 0$.
In the first case the current $I_{sc}$ and the electric field $E =
-\bigtriangledown V - (1/c)dA/dt =  -\bigtriangledown V - (1/cl)d\Phi/dt$
have the same direction in all segments, i.e. each segment $l_{s}$ is a load in which
the power $\int_{l_{s}}dl I_{sc}E$ is dissipated. In the second case the persistent
current and the electric field $\overline{E} = -\overline{\bigtriangledown V}$
should have opposite directions in one of the segments because $\int_{l}dl
\bigtriangledown V \equiv 0$, i.e. this segment is a dc power source
$W = V_{os}I_{p.c.}$ at $V_{os} \neq 0$.

Already the classical LP experiment is evidence of the dc power source because
an energy dissipation with power $R_{l}I_{p.c.}^{2}$ takes place at $R_{l} > 0$
and $I_{p.c.} \neq 0$. The conventional current at $R_{l} > 0$ is maintained by
the Faraday's voltage. In the case of  the persistent current the Faraday's voltage
is substituted by the quantum force \cite{QuaForce}. According to \cite{QuaForce}
the quantum force as well as the Faraday's voltage is distributed uniformly among
the loop. This theoretical result is corroborated by our observation of the
voltage oscillations Figs.3,4 described by the analogy with a conventional loop (4).

\begin{figure}[bhb]
\vspace{0.1cm}\hspace{-1.5cm}
\vbox{\hfil\epsfig{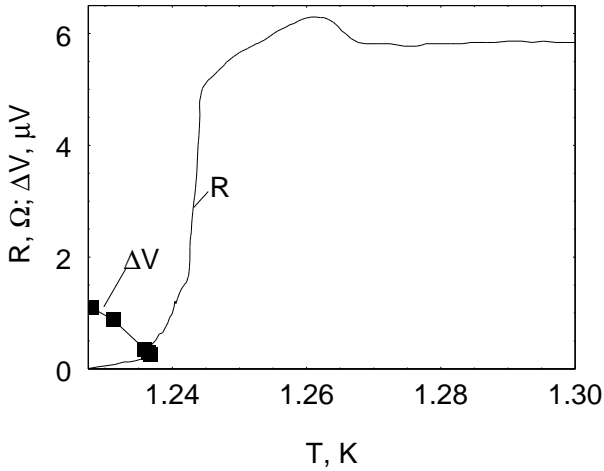}\hfil}
\vspace{0.75cm}
\caption{The temperature dependence of the amplitude  $\Delta V $ of the voltage
oscillations on the $V_{2}$ contacts of the asymmetric loop
with 2r = 4 $\mu m$ and w = 0.4 $\mu m$ relatively of its resistive transition R measured at $I_{m} = 1.5 \ \mu A$. } \label{fig-1}
\end{figure}

According to \cite{QuaForce} the voltage Figs.3,4  observed at $I_{m} = 0$
is induced by a switching of the loop between superconducting states
with different connectivity, i.e. between the states with $<n_{s}^{-1}>^{-1} = 0$
and $<n_{s}^{-1}>^{-1} \neq 0$ and its value is proportional to the average
frequency of the switching $\Delta V \propto \omega_{sw}$. This switching
at $T \approx T_{c}$ can be induced  by both thermal fluctuations and an external
electric noise. Therefore the chaotic energy of thermal fluctuations or an external
electric noise is transformed in the dc power $W = V_{os}I_{p.c.}$ according to \cite{QuaForce} .

The amplitude $\Delta V$ of the voltage oscillation induced by thermal fluctuations
should have a maximum value near $T_{c}$ because the frequency $\omega_{sw}$
of the switching induced by the fluctuations is maximum near $T_{c}$.
We observed the oscillations in the temperature region corresponds
to the bottom of the resistive transition Fig.5. The amplitude $\Delta V $ of the oscillation increases
with temperature lowering down to the lowest  temperature we could reach, Fig.5.
At higher temperature the signal/noise ratio is not adequate because of small $\Delta V $ value.

The temperature dependence  $\Delta V (T)$ Fig.5 shows that the
voltage oscillations Fig3,4 are induced rather by
a high-frequency external electric noise. The frequency $\omega_{sw}$
of the switching induced by the electric noise
can faintly decrease with the temperature lowering and the $\Delta V $
value should increase because $I_{p.c.} \propto T_{c} - T$.

We assume that the dc voltage induced by the fluctuations is smaller
than the value which we can detected. The
limit value of the power which can be induced by the thermal fluctuations
$(k_{B}T)^{2}/\hbar \approx 10^{-12} \ Wt$ \cite{QuaForce}. We detected
the power down to $10^{-14} \ Wt$. But in our real case the power induced
by the fluctuations can be much lower than the theoretical limit and lower than
we can detected. We observed in some samples voltage oscillations at the
temperature corresponded to the top of the resistive transition where any
electric noise should not influence on the voltage value. But this observation
can not be considered as an evidence of the power induced by the fluctuation
because it is no enough reliable.

Although our experimental resources did not allow to detect the dc voltage
induced  by thermal fluctuations our result is new in essence. It can not be
explained by conventional rectification of an external electric noise which was
observed in superconductors. Such rectification is explained by an asymmetry
of the sample but it can not explain the voltage oscillations Figs.3,4. Only
reasonable and natural explanation of these oscillations is the relation (4)
followed from \cite{QuaForce}. In this relation $j_{p.c.} \propto (n - \Phi/\Phi_{0})_{min}$
because of the quantization and $<\rho >_{l_{s}} > 0$ because of reiterated switching
of $l_{s}$ in the state with $\rho > 0$.

In conclusion, we have observed voltage oscillations measured on segments
of an inhomogeneous loop at zero external direct current. This result corroborates
the theoretical prediction  \cite{QuaForce} according to which such voltage
oscillations can be induced by thermal fluctuations or an external
electric noise. We can detect the dc voltage induced only by
an external electric noise.

We acknowledge useful discussions with V.A.Tulin and thank A.M.Orlov
for technical support.

\end{document}